\documentclass[10pt,showpacs,twocolumn]{revtex4}
\usepackage{color}
\usepackage{graphicx}
\usepackage{amsmath}
\usepackage{amssymb}
\usepackage{}
\begin{document}

\preprint{APS/123-QED}


\title{Spectrally resolved spatiotemporal features of quantum paths in high-order harmonic generation}

\author{Lixin He, Pengfei Lan,\footnote{pengfeilan@mail.hust.edu.cn} Qingbin Zhang, Chunyang Zhai, Feng Wang, Wenjing Shi, and Peixiang Lu\footnote{lupeixiang@mail.hust.edu.cn}
}

\affiliation{%
 School of Physics and Wuhan National Laboratory for Optoelectronics, Huazhong University of Science and Technology, Wuhan 430074, China\\
}%

\date{\today}

\begin{abstract}
We experimentally disentangle the contributions of different quantum paths in high-order harmonic generation (HHG) from the spectrally and spatially resolved harmonic spectra.
By adjusting the laser intensity and focusing position, we simultaneously observe the spectrum splitting, frequency shift and intensity-dependent modulation of harmonic yields both for the short and long paths. Based on the simulations, we discriminate the physical mechanisms of the intensity-dependent modulation of HHG due to the quantum path interference and macroscopic interference effects. Moreover, it is shown that the atomic dipole phases of different quantum paths are encoded in the frequency shift. In turn, it enables us to retrieve the atomic dipole phases and the temporal chirps of different quantum paths from the measured harmonic spectra. This result gives an informative mapping of spatiotemporal and spectral features of quantum paths in HHG.
\end{abstract}                         
\maketitle

\section{Introduction}

Nowadays, high-order harmonic generation (HHG) through the interaction of intense laser pulses with atomic or molecular gases has been extensively investigated to produce coherent extreme ultraviolet (XUV) radiations \cite{zc,eco,js} and attosecond pulses \cite{hent,pm,eg,zk,ej}.
These ultrashort pulses can serve as an important tool for detecting
the ultrafast electron dynamics inside atoms or molecules \cite{rk,zy,mu} as well as inaugurating a new domain for time-resolved metrology and spectroscopy on attosecond time scale \cite{hent,md,atto}. Due to these applications, HHG has been a subject of great interest in the past two decades.

HHG process includes both the response of the single atom exposed to laser field and also the macroscopic copropagation of the laser field and high harmonics.
The physics underlying single-atom response can be well understood by the three-step model \cite{Threestep}: (i) ionization, (ii) acceleration and (iii) recombination. Following this idea, a quantum theory has also been developed within the strong-field approximation (SFA) \cite{Lewenstein}, which also can be incorporated into the framework of Feynman's path
integral theory \cite{fman}. In this approach, the harmonic dipole moment can be written as the coherent sum over all different quantum paths contributing to HHG. These quantum
paths are a generalization of the classical electron trajectories. For each harmonic in the plateau region, two families of electron trajectories with the shortest excursion times ($\tau$) in the continuum, which are commonly referred to the short and
long quantum paths, dominantly contribute to the harmonic emission. The phase associated to each quantum path of the $q$th harmonic emission is given by the action of the electron accumulated in the laser field, which is usually called atomic dipole phase and can be approximated by $\Phi_{q}^{j}\approx-U_p\tau_{q}^{j}\approx-\alpha_{q}^{j}I$ in terms of the SFA model \cite{Lewenstein2}. Here, $U_p$ and $I$ are the ponderomotive energy and laser intensity and $\alpha_{q}^{j}$ is the phase coefficient with $j=S,L$ representing the short and long quantum paths, respectively.

The single-atom dipole phase is naturally coupled with the macroscopic response, which leads to different phase-matching and therefore rich spatial and spectral features of HHG \cite{Dubrouil}. Thus, in turn, the spatial and spectral features offer useful information on the HHG process \cite{Bellini,Mansten} and the gas medium \cite{Itatani}. Moreover, each path has a specific excursion time and frequency property and it leads to a mapping between time and frequency for each harmonic \cite{Sansone} which has also been used to detect the nuclear dynamics \cite{Baker}. On the other hand, the quantum paths can interfere with each other, resulting in frequency splitting \cite{Kan,Wang,Zhong,wei,Brunetti} and interference fringes \cite{HeXK,LiuPeng,hx} in the harmonic spectra and also laser intensity-dependent interference of harmonic yield \cite{Zair,Heyl,fs,holl,azc,tau,dao}. The interference fringes allow to retrieve even
more information about electron motion
with high precision. However, because the single-atom and macroscopic phase-matching effects are naturally coupled in HHG process, the mechanisms underlying the interference and spectral features are always entangled and sometimes the interpretation is different or confusing even for the similar observations \cite{Kan, Brunetti,HeXK,LiuPeng,hx,Zair,Heyl}.
Experimental investigations of fully resolved HHG with both spatiotemporal and spectral features will shed more light on these issues. It is of crucial importance, both from a
fundamental point of view, to understand the laser-atom interaction, as well as for applications such as HHG spectroscopy \cite{npy}. However, because the phase coefficient of the long path is almost one-order of magnitude higher than that of the short path, the phase-matching is quite different and it becomes very challenging to
simultaneously resolve the spatiotemporal and spectral features and interference fringe both for the long and short paths.

\begin{figure}[htb]
\centerline{
\includegraphics[width=8cm]{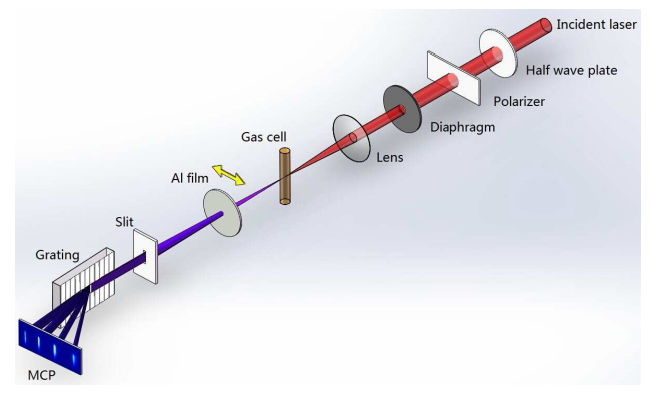}}
 \caption{\label{fig1}(Color online) Experimental setup used for the harmonic generation and detection.
 }
\end{figure}

In this work, we aim at comprehensively disentangling the HHG features of different quantum paths with spectrally and spatially resolved harmonic spectra and identifying the mechanisms of the intensity-dependent interference of HHG. In our experiment, we observe that each individual high-order harmonic is spectrally split, i.e., shows a double-peak structure. Those two peaks are spatially separated in the far field and each peak also presents different frequency shift as increasing the laser intensity. By adjusting the laser intensity and focusing position, each peak shows different spatial profiles, either blue- or red- frequency shift and intensity-dependent periodic modulation. By performing the simulations of both the single-atom and macroscopic responses of HHG, we differentiate the intensity-dependent interferences caused by the quantum path interference (QPI) and the transient phase matching. In comparison with simulation and experiment, the measured spectral and spatial features are well explained. To the best of our knowledge, it is the first time to observe the spectrum splitting, frequency shift and intensity-dependent interference simultaneously for the short and long quantum paths in experiment. Moreover, we show that the atomic dipole phases of different quantum paths are encoded in the frequency shift. In turn, it enables us to retrieve the atomic dipole phases and the temporal chirps of different quantum paths from the measured harmonic spectra.
\begin{figure}[htb]
\centerline{
\includegraphics[width=8cm]{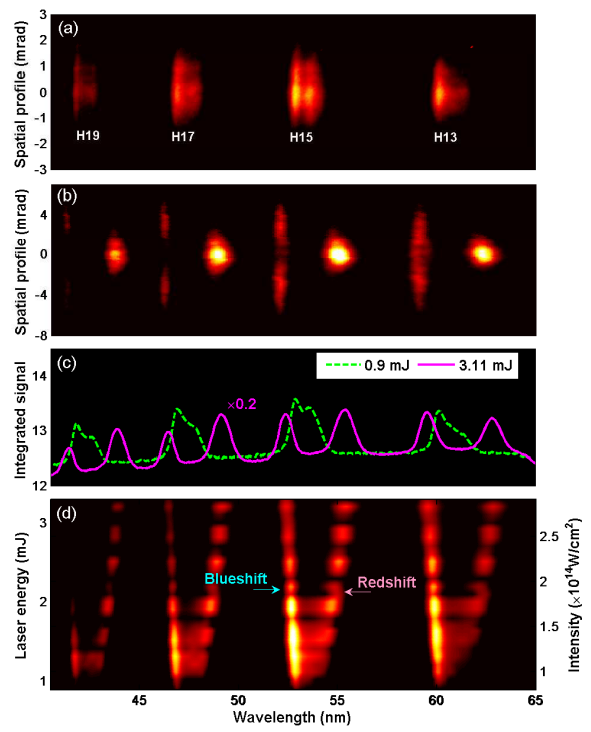}}
 \caption{\label{fig2}(Color online) (a)-(b) Measured harmonic spectra at the laser energies of 0.9 mJ and 3.11 mJ, respectively. (c) The spatially integrated HHG signals for the spectra in (a) (dashed line) and (b) (solid line). For the purpose of clarity, the solid line is multiplied by a factor of 0.2. (d) Spatially integrated HHG signals as a function of the laser intensity. Here, the gas cell is located at Z=2 mm.
 }
\end{figure}

\begin{figure*}[htb]
\centerline{
\includegraphics[width=13cm]{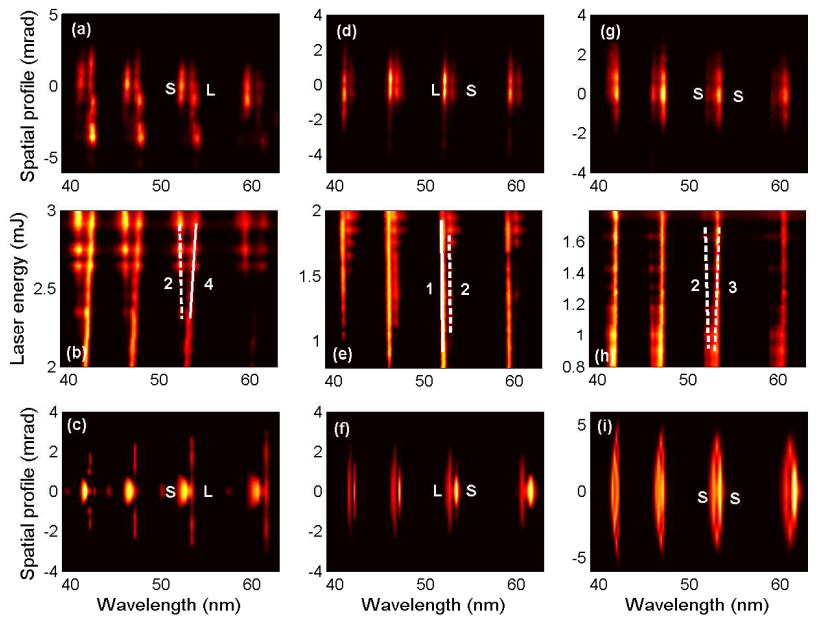}}
 \caption{\label{fig3} (Color online) (a) Measured harmonic spectrum with the gas cell located at Z=$-$3 mm, (b) spatially integrated harmonic spectra as a function of the laser energy, and (c) calculated spatially resolved harmonic spectrum to compare with (a). (d)-(f) and (g)-(i) are the same to (a)-(c), but for the gas cell located at Z=0 mm and Z=3 mm, respectively.
 }
\end{figure*}

\section{Spatially and spectrally resolved high order harmonics}
Our experiment is carried out using a commercial Ti:sapphire laser system (Legend Elite-Duo, Coherent, Inc.), which delivers the 30-fs, 800-nm pulses at a repetition rate of 1kHz with a maximum pulse energy of 10 mJ. Figure \ref{fig1} shows the sketch of the experimental setup. In brief, the incident laser beam is focused to a 2-mm long gas cell filled with argon with a pressure of 35 torr by a 600-mm focal-length lens. The gas cell is mounted on a three axis picomotor actuator controlled by a computer. An argon gas jet emitted from a 500-$\mu$m
diameter nozzle has also been tried and similar phenomena (spectrum splitting and frequency shift) can also be observed. Of course, by using a gas jet, the experimental conditions (e.g., the gas pressure and interaction distance) are different from that with a gas cell. A half-wave plate and a polarizer are placed in the light path to continuously control the laser
energy from 0.1 mJ to 5 mJ, and an iris diaphragm is used to adjust the laser beam size (around 9-mm diameter in the experiment). A 500-nm thick aluminum foil is placed in front of the spectrometer to block the driving pulse.
The generated harmonic spectrum is detected by a home-made flat-field soft x-ray spectrometer consisting of a flat-field grating (1200 grooves/mm) and a slit with a width of about 0.1 mm and hight of 15 mm. High order harmonics are dispersed by the grating and imaged onto the Micro-channel plate (MCP) fitted with a phosphor screen. The image on the screen is read out by a charge-coupled device (CCD) camera.

Figures \ref{fig2}(a) and \ref{fig2}(b) display the spatially resolved harmonic spectra obtained at the laser energies of 0.9 mJ and 3.11 mJ respectively,
and their corresponding spatially integrated HHG signals are presented as the dashed (0.9 mJ) and solid (3.11 mJ) lines in Fig. 2(c).
In our experiment, we have estimated the laser intensity by measuring the laser power and the focus size. The laser power can be directly read out from a power meter. The focus size is measured by using a CCD beam profiler that mounted on a translation stage in the direction of the beam propagation, and finally the measured focal spot is demonstrated to have a diameter about 200 $\mu$m. Due to the dispersion of the optical elements (e.g., a 3-mm thick window plate in the front of the vacuum chamber and a 3-mm thick half-wave plate), the 30-fs fundamental laser pulse delivered by the laser system will be broadened to about 36 fs. Finally, the laser intensity is estimated about $2.75\times10^{14}\ \mathrm{W/cm}^2$ and $0.8\times10^{14}\ \mathrm{W/cm}^2$ at the laser energy of 3.3 mJ and 0.9 mJ, respectively. However, with this method, the uncertainty of the estimated laser intensity may be up to 20\%. The laser intensity can also be determined from the cutoff of the measured harmonic spectra. However, due to the limitation of the dynamic range of CCD, the cutoff position is difficult to identify from the generated harmonic spectra. The gas cell is placed at Z$=2$ mm. Here Z$<0$ and $>0$ means the upstream and downstream of the laser focus, respectively. One can clearly see several remarks from Fig. \ref{fig2}.
First, each individual harmonic is split into two branches both in the spatial and the spectral domains. The spatial and spectral separations at high laser intensity (3.11 mJ, $\sim2.75\times10^{14}\ \mathrm{W/cm}^2$) are much
larger than those at a low laser intensity (0.9 mJ, $\sim0.8\times10^{14}\ \mathrm{W/cm}^2$). Second, the spatial divergence of left branch is much broader ($\pm$6 mrad) and depends more sensitively on the laser intensity
than that of the right one ($\pm$2 mrad). This suggests us to assign the left branch as the long path contribution and the right one as the short path contribution,
since the long path requires a lager divergence angle for phase matching than the short path \cite{ppm}. Third, as shown in Fig. \ref{fig2}(d), the left branch of each individual harmonic (long path)
is clearly blue-shifted, while the right branch (short path) is red-shifted in the frequency domain as the laser intensity increases.
For simple description, in the following, we abbreviate the blue and red shift of the short (long) path as SB (LB) and SR (LR) shift, respectively.

\begin{figure}[htb]
\centerline{
\includegraphics[width=8.5cm]{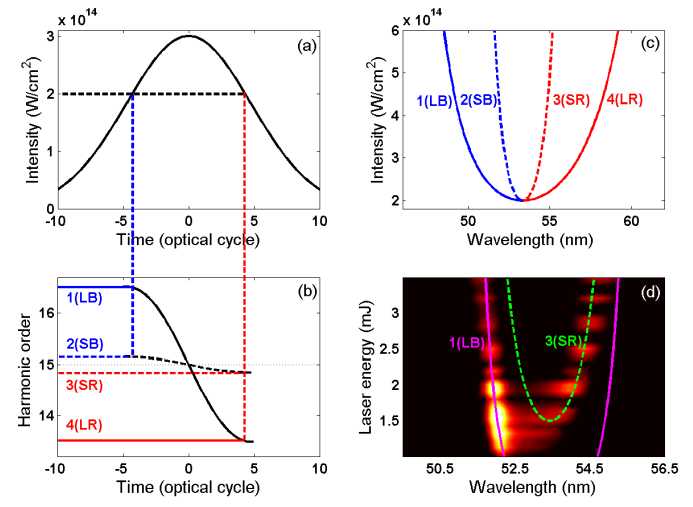}}
 \caption{\label{fig4}(Color online) (a) Temporal profile of the laser intensity. (b) Instantaneous frequencies of H15 for the short (dashed line) and long (solid line) paths. (c) Instantaneous frequencies of H15 as a function of the laser intensity. (d) Comparison between experiment and simulation for the intensity-dependent 15th harmonic emission.
 }
\end{figure}

In Fig. \ref{fig3}, we have presented the harmonic emission at different focusing geometries. Figure \ref{fig3}(a) shows the spatially resolved harmonic spectrum generated with the laser intensity of $2.3\times10^{14}\ \mathrm{W/cm}^2$ (2.6 mJ) at Z=$-$3 mm. Figure \ref{fig3}(b) shows the spatially integrated harmonic spectra as a function of laser intensity. All these individual harmonics are spatially and spectrally separated. However, the spatial and spectral profiles in Figs. \ref{fig3}(a) and (b) are quite different from Fig. \ref{fig2}. One can see that, at Z=$-$3 mm [Figs. \ref{fig3}(a), (b)], the left branch is distributed in a narrower space range ($\pm$2 mrad) and blue-shifted in the frequency domain. While the right branch presents a broader spatial distribution within $-5$$\sim$3 mrad and a red-shift. In other words, the left and right branches in Figs. \ref{fig3}(a) and (b) show the SB and LR shift. In contrast, those in
Fig. \ref{fig2} show the LB and SR shift.
Distinct and interesting spatial and spectral features also can be observed at Z=0 mm [Figs. \ref{fig3} (d) and (e)] and Z=3 mm [Figs. \ref{fig3}(g) and (h)]. To be specific, for Z=0 mm, both the left and right branches are blue-shifted. However, the left branch shows a large spatial divergence (i.e., LB) and the right branch shows a small spatial divergence (i.e., SB). In contrast, for Z=3 mm, each of the two branches has a narrow spatial divergence ($\pm$2 mrad). However, the left branch is blue-shifted (i.e., SB) and the right branch is red-shifted (i.e., SR). In addition, we have also calculated far-field spatial profiles of harmonics at Z=$-$3 mm, Z=0 mm, and Z=3 mm to compare with the experimental results in Figs. \ref{fig3} (a), (d), and (g). Corresponding results are presented in Figs. \ref{fig3} (c), (f), and (i), respectively. One can see that, the calculated results agree reasonablely with the measurements: each harmonic is split into two branches, and the spatial distributions of the separated two branches (attributed to short or long path) are in reasonable agreement with the experiments. Some differences in details are likely due to the uncertainties of the experimental conditions, e.g., the laser intensity and focus size. In short, by adjusting the laser intensity and focusing position, we have simultaneously observed rich spatio-spectral features of the spectrum splitting and frequency shift both for the short and long paths, which have not been observed so far.

The observed spectrum splitting and frequency shift can be explained by considering the macroscopic phase-matching of HHG \cite{ppm,cgd,tp}.
The phase mismatch between the driving laser field and the emitted high order harmonics can be expressed by $\Delta k=\Delta k_q+\Delta k_g+\Delta k_a+\Delta k_e$.  Here $\Delta k_q$ is the phase mismatch due to the single-atom dipole phase and can be written as $-\alpha_q \nabla I$. $\Delta k_g$ is the geometrical wave vector mismatch due to focusing. $\Delta k_a$ and $\Delta k_e$ are the wave vector mismatches between high harmonics and fundamental field due to the dispersion in the neutral medium and free electrons. Due to the variation of laser intensity with time, $\Delta k_q$ and therefore $\Delta k$ will change during the laser pulse.
This leads to transient phase matching of different quantum paths. The variation of the instantaneous frequency of each harmonic can be expressed by \cite{Heyl,mbg,ps}
\begin{eqnarray}
\omega_q(t)=q\omega_0+\alpha_{q}^{j}\frac{\partial I(t)}{\partial t}.
\end{eqnarray}
In Fig. \ref{fig4}(b), we show the instantaneous frequencies of the short (dashed line) and long (solid line) paths for the 15th harmonic (H15) as an example.
One can see that at the uphill of the laser pulse, both the short and long paths present a blue shift, i.e., the SB and LB shift. While at the downhill,
red shift is presented for these two paths, i.e., the SR and LR shift. Since the phase coefficient $\alpha_{q}^{S} \ll \alpha_{q}^{L}$,
the frequency shift of the short path is much weaker than the long one, thus is more difficult to be observed in experiments.
During the laser pulse, the phase matching can be transiently optimized at certain intensities $I_m$ due to the intensity dependence of the atomic dipole phase. By inserting a Gaussian intensity envelope (i.e., $I(t)=I_0exp(-4\ln2 t^2/\tau^2)$, where $I_0$ and $\tau$ denote the peak intensity and the pulse duration full width at half maximum) into Equation (1), one can then calculate the $I_0$-dependent instantaneous harmonic frequency at a certain phase matching intensity $I_m$ by \cite{Heyl,marker2}:
\begin{eqnarray}
\omega_q(I_0)=q\omega_0\pm \alpha_{q}^{j}\frac{4\sqrt{\ln2}I_m}{\tau}\sqrt{\ln(I_0/I_m)}.
\end{eqnarray}
Figure \ref{fig4}(c) shows the calculated intensity dependence of the instantaneous frequency of H15 with $I_m$=$2.0\times10^{14}\ \mathrm{W/cm}^2$. With the increase of the laser intensity, H15 is gradually separated into four branches, as marked as the 1(LB), 2(SB), 3(SR), and 4(LR), respectively.
When compared with the experimental result at Z=2 mm, we find that evolutions of the measured two branches match well with the branch 1 (LB, left) calculated with $I_m$=$0.23\times10^{14}\ \mathrm{W/cm}^2$ (0.26 mJ) and branch 3 (SR, right) with $I_m$=$1.34\times10^{14}\ \mathrm{W/cm}^2$ (1.52 mJ), respectively. In our experiment, we have observed various combinations of the four branches at different focusing positions. However, it is very challenging to capture these four branches simultaneously, because the phase coefficient of the long path is almost one-order of magnitude higher than that of the short path.

\begin{figure}[htb]
\centerline{
\includegraphics[width=8cm]{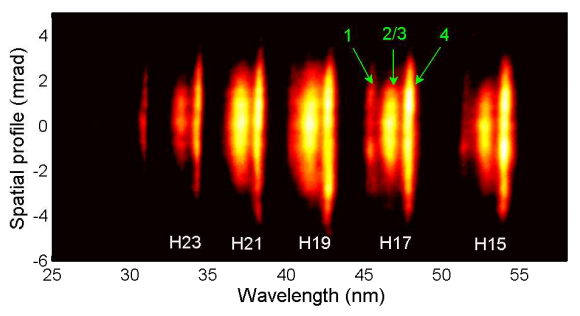}}
 \caption{\label{fig5}(Color online) The harmonic spectrum measured at the intensity of $2.9\times10^{14}\ \mathrm{W/cm}^2$ (3.3 mJ) with the gas cell placed at Z=$-$6 mm.
 }
\end{figure}

Tracing the four branches requires strict experimental conditions to satisfy the phase matching of both long and short paths and also large frequency splitting. To this end, we have tried to observe the harmonic spectrum at Z=$-$6 mm (to ensure the coexist of the short and long paths) with a high intensity about $2.9\times10^{14}\ \mathrm{W/cm}^2$ (3.3 mJ, to make the two paths both well separated). Under this condition, the ionization probability is so high that the harmonic yield decreases rapidly due to the dispersion induced by free electrons and the absorption in the gas medium.
Figure \ref{fig5} shows the spatially resolved harmonic spectrum. One can see that the 15th and 17th harmonics have been split into 3 (or 4) branches in the spatial and spectral domains. The outer most two branches exhibit relatively broader spatial profiles compared to the inner one, which are mainly due to the separation of the long path, corresponding to the branches 1 and 4, respectively. While the inner one is the coalition of the branches 2 and 3 from the short path.

\begin{figure}[htb]
\centerline{
\includegraphics[width=8cm]{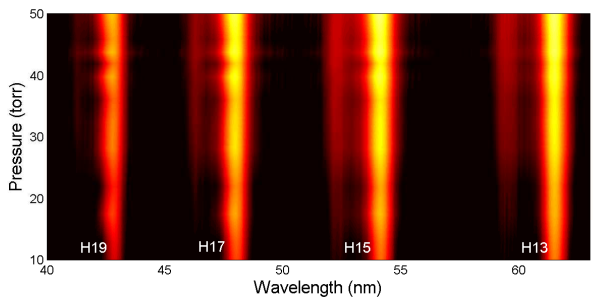}}
 \caption{\label{fig6}(Color online) Spatially integrated HHG signals as a function of the gas pressure. Here, the gas cell is located at Z=2 mm, and the laser intensity is fixed at $1.77\times10^{14}\ \mathrm{W/cm}^2$ (2 mJ).
 }
\end{figure}

In our experiment, the phase matching conditions of different quantum paths are changed by adjusting the laser intensity and focusing position, which mainly affects the values of $\Delta k_q$ and $\Delta k_g$ and therefore changes the $\Delta k$. When the laser intensity is high enough, large numbers of electrons are ionized. The mismatching due to the dispersion of free electrons ($\Delta k_e$) as well as the depletion of the ground state becomes serious and cannot be neglected, which will consequently lead to a rapid decrease in the harmonic yield (as mentioned in Fig. \ref{fig5} above). However, in this case, the spectrum splitting can still be observed from the measured harmonic signals (as shown in Fig. \ref{fig5}). For the dispersion in the
neutral medium $\Delta k_a$, it depends on the gas pressure. In Fig. \ref{fig6}, we show the measured harmonic spectra as a function of the gas pressure. Here, the gas cell is located at Z=2 mm, and the laser intensity is fixed at $1.77\times10^{14}\ \mathrm{W/cm}^2$ (2 mJ). With the gas pressure changing from 10 to 50 torr, the spectrum splitting of each harmonic can always be observed, but the strengths of the separated two branches increase gradually. These results indicate that the ionization and dispersion of the
gas medium ($\Delta k_e$ and $\Delta k_a$) could have an effect on the measured harmonic strength, but will not affect the spectrum splitting of the generated harmonics.

On the other hand, it should be noted that the separated branches in harmonic spectra show obvious asymmetry.
For instance, in Fig. \ref{fig4} (d), the harmonic emission from 1LB is stronger than that from 3SR, especially in the intensity range from $0.88\times10^{14}\ \mathrm{W/cm}^2$ (1 mJ) to $1.77\times10^{14}\ \mathrm{W/cm}^2$ (2 mJ).
And in Fig. \ref{fig5}, the branch 4 (LR) is much brighter than branch 1 (LB) and dominate the harmonic emission.
To clarify this asymmetry of the harmonic splitting, we next investigate the phase matching of these separated branches.
The phase mismatch $\Delta k$ for both the short and long paths can be calculated with the theory in \cite{ppm}.
In the following, we take the 15-th harmonic as an example. Under our experimental condition, the refractive indexes of the laser field and harmonics in the gas are very close and therefore the phase mismatching due to the dispersion in the neutral medium ($\Delta k_a$) is very small and can be neglected.
In the case of Fig. \ref{fig4} (d), the dominant HHG contributions are from the long path with a blue shift (1LB) and the short path with a red shift (3SR).
Then we analysis the phase matching of the long (short) path at the uphill (downhill) of the laser pulse.
The laser intensity here is chosen as $1.5\times10^{14}\ \mathrm{W/cm}^2$ to ensure the asymmetry is distinct. Under this intensity,
the ionization probability of electrons is below $2\%$, the electronic dispersion ($\Delta k_e$) is small and can also be neglected.
We have calculated the phase mismatching $\Delta k_q$ and $\Delta k_g$ for the long path at $t=-10$ fs (uphill) and the short path at $t=10$ fs (downhill).
The result shows that, for the long path, $\Delta k_q$=$-$264.6/m, $\Delta k_g$=756.1/m, and the total phase mismatching $\Delta k$=491.5/m.
While for the short path, $\Delta k_q$=$-$16.5/m, $\Delta k_g$=756.1/m, and $\Delta k$=739.6/m. One can clearly see that for the short and long paths,
$\Delta k_g$ is the same, but $\Delta k_q$ is quite different. Finally£¬the total phase mismatching of the long path is smaller than that of the short one.
As a consequence, the harmonic emission from the branch 1LB is stronger than 3SR.
This result indicates that the asymmetry of the two branches in Fig. \ref{fig4} (d) is mainly due to the different single-atom dipole phases of the short and long paths.
Next we consider the phase matching in a more intense laser field, e.g., $2.9\times10^{14}\ \mathrm{W/cm}^2$ as in Fig. \ref{fig5}.
In this case, the ionization probability is up to $50\%$. The electronic dispersion term ($\Delta k_e$) can no longer be ignored.
We have calculated the phase mismatching $\Delta k_q$, $\Delta k_g$, and $\Delta k_e$ for the long path at t=$-$10 fs (uphill, for branch 1LB) and t=10 fs (downhill, for branch 4LR).
For branch 1LB, $\Delta k_q$=1310/m, $\Delta k_g$=698.7/m, and $\Delta k_e$=$-$6.5/m. While for 4LR, $\Delta k_q$=1310/m, $\Delta k_g$=698.7/m, and $\Delta k_e$=$-$1050/m.
One can see that both $\Delta k_q$ and $\Delta k_g$ are equal for these two branches. But the electronic dispersion ($\Delta k_e$) becomes more serious at t=10 fs.
This is due to the larger density of free electrons at the downhill of the laser pulse. The total phase mismatching $\Delta k$ of branch 1LB is 2002.2/m, whereas it is only 958.7/m for branch 4LR.
In other words, the electronic dispersion $\Delta k_e$ can compensate the mismatching due to the dipole phase and geometrical focusing, the total mismatching of the long path is smaller at the downhill of the laser pulse. Therefore, the branch 4LR is brighter than the branch 1LB and dominate the harmonic emission.

\section{Intensity-dependent interference of HHG yield}

\begin{figure}[htb]
\centerline{
\includegraphics[width=8.5cm]{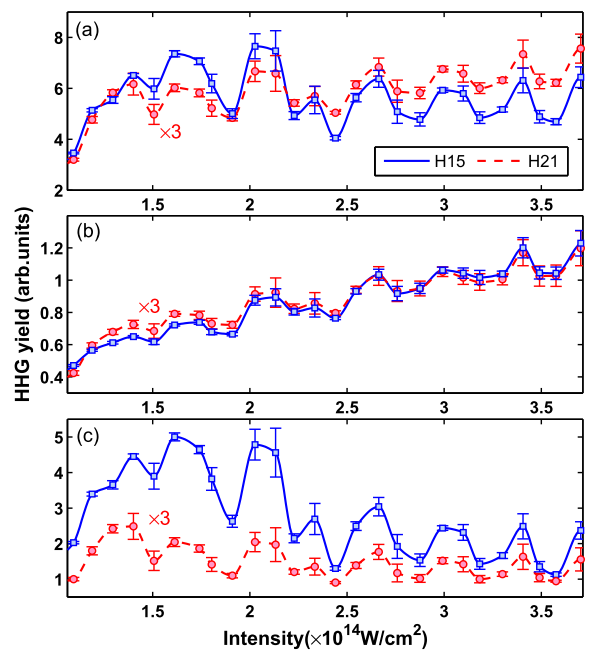}}
 \caption{\label{fig7}(Color online) Spatially integrated harmonic signals of H15 and H21 generated at Z=2 mm with (a) no spatial filtering, (b) off-axis spatial filtering (from 2 to 6 mrad), and (c) on-axis spatial filtering (from $-$2 to 2 mrad). For a better comparison, the dashed line is multiplied by a factor of 3.
 }
\end{figure}

Apart from the spectrum splitting and the frequency shift discussed above, one can also observe pronounced periodic modulations of HHG yields from Fig. \ref{fig2}(d). In detail, we have presented in Fig. \ref{fig7} the intensity dependence of HHG yields for H15 (solid lines) and H21 (dashed lines) with the gas cell placed at Z=2 mm. Figure \ref{fig7}(a) shows the result obtained without any spatial filtering, where both the short and long paths are included. As shown in this figure, these two harmonics show the same modulation periodicity of about $0.3\sim0.4\times10^{14}\ \mathrm{W/cm}^2$. In Fig. \ref{fig7}(b), the off-axis spatial filtering (from 2 to 6 mrad) is performed such that the long path is selected. The obtained modulations are basically the same as in Fig. \ref{fig7}(a), but the modulation amplitude is reduced. Similar results are also found in Fig. \ref{fig7}(c), where a spatial filter on axis (from $-$2 to 2 mrad) is used such that the short path dominates the harmonic emission. These results indicate that the modulation of HHG yields in our experiment depends insensitively on the harmonic order as well as the dominant quantum path.

\begin{figure}[htb]
\centerline{
\includegraphics[width=8cm]{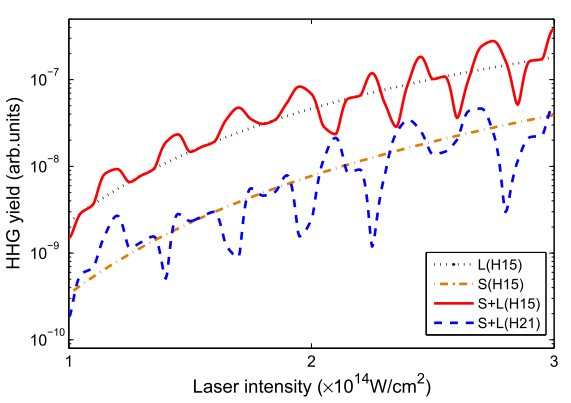}}
 \caption{\label{fig8}(Color online) Calculated HHG yields with the single-atom response. Solid and dashed lines are for H15 and H21 with both the short and long paths contributions. The dotted and dash-dotted lines are for the 15th harmonic with only the long or short path contribution. Here, S and L represent the short and long paths.
 }
\end{figure}

\begin{figure*}[htb]
\centerline{
\includegraphics[width=13cm]{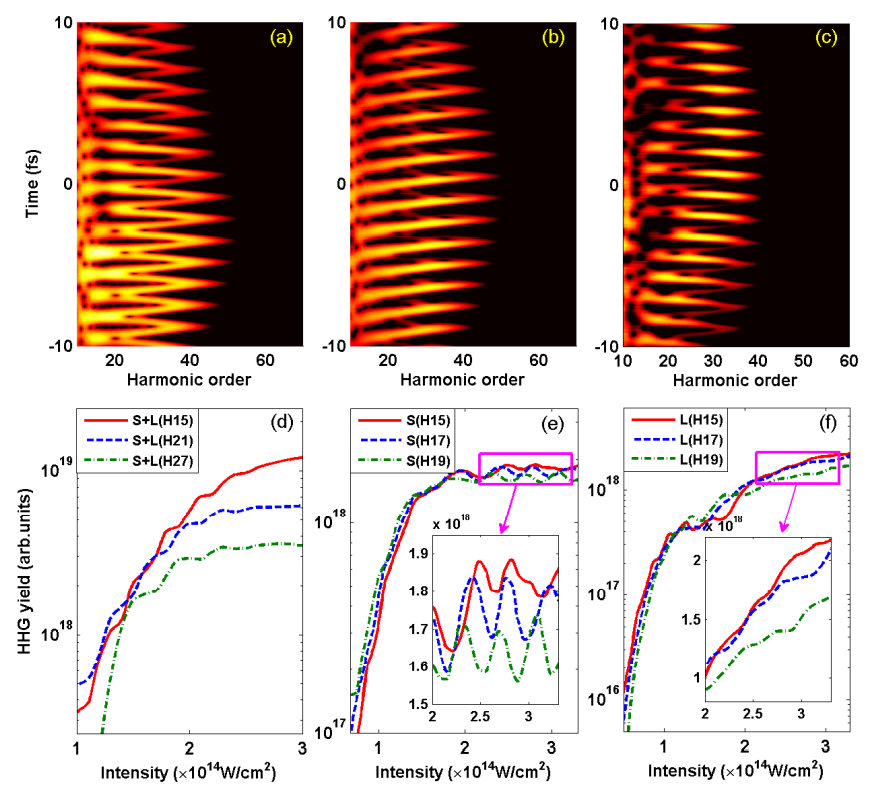}}
 \caption{\label{fig9}(Color online) (a)-(c) Time-frequency spectrogram of the harmonic spectra generated at $2.5\times10^{14}\ \mathrm{W/cm}^2$ by putting the gas cell at the focus (Z=0 mm), the downstream (Z=2.5 mm), and the upstream (Z=$-$2.5 mm).
 (d)-(f) Intensity dependence of HHG yields calculated for different harmonic orders.
 }
\end{figure*}

Similar intensity-dependent modulations of HHG yields have been previously reported \cite{Heyl,Zair}. However, the interpretation is different and still confusing. At the single-atom level, the modulation has been attributed to the QPI between the short and long paths due to their different intensity-dependent
dipole phases \cite{Zair}. While on the macroscopic aspect, the transient phase matching due to the variation of $\Delta k$ with time can also lead to oscillation of HHG yield, which is also called temporal Marker fringes \cite{Heyl,marker1,marker2}. To disentangle these two physical mechanisms, we perform the simulations of harmonic emission based on both the single-atom and macroscopic phase-matching responses. Figure \ref{fig8} shows the calculated HHG yields with the single-atom response \cite{Zair}.
It is obvious that when the short and long paths coexist, QPI occurs and induces distinct modulations in the intensity-dependent HHG yields for both H15 (solid line) and H21 (dashed line). Whereas, the modulation periods for these two harmonics are slightly different (about $0.25\times10^{14}\ \mathrm{W/cm}^2$ for H15 and $0.3\times10^{14}\ \mathrm{W/cm}^2$ for H21). This is caused by the different $\Delta\alpha_q$ (difference of the two phase coefficients, i.e., $\alpha_{q}^{L}-\alpha_{q}^{S}$) of H15 and H21 and the modulation period is approximately equal to $2\pi/\Delta\alpha_q$. In Fig. \ref{fig8}, the intensity-dependent HHG yield of H15 with only the long (short) path contribution is also presented as the dotted (dash-dotted) line. With only one quantum path contribution, QPI no longer occurs and therefore the periodic modulation disappears. This simulation shows that the intensity-dependent interference caused by QPI demands the coexistence of the short and long paths and the interference fringe depends on the harmonic order.

The macroscopic response in the gas medium is described by the copropagation of the laser and high harmonic fields, which is simulated by numerically solving
the Maxwell wave equations as in \cite{pro,pro_lan}.
Similar to our experiment, the gas medium position is changed in the simulations to satisfy the phase-matching of the short, long and both quantum paths.
Figs. \ref{fig9}(a)-\ref{fig9}(c) display the time-frequency spectrogram calculated by the Gabor transform \cite{Gabor}, of the generated harmonic spectra at $2.5\times10^{14}\ \mathrm{W/cm}^2$ by putting the gas cell at the focus (Z=0 mm), the downstream (Z=2.5 mm), and the upstream (Z=$-$2.5 mm), respectively. One can clearly see that, by putting the gas cell at the focus, both the short and long paths [corresponding to the lower and upper branches of each peak in Fig. \ref{fig9}(a)] exist in each half optical cycle. While at the downstream, only the short (lower branch) path is dominant. While at the upstream, the long path (upper branch) is well selected. We have calculated the HHG yields (spatially integrated in the far field) for different harmonic orders at the above three focusing positions. Corresponding results are shown in Figs. \ref{fig9}(d)-\ref{fig9}(f), respectively. As shown in Fig. \ref{fig9}(d), due to the QPI of the short and long paths, one can clearly see the intensity-dependent modulations of HHG yields. The modulation period depends on the harmonic order. It is about $0.25\times10^{14}\ \mathrm{W/cm}^2$ for H15, $0.3\times10^{14}\ \mathrm{W/cm}^2$ for H21, and $0.4\times10^{14}\ \mathrm{W/cm}^2$ for H27. In Fig. \ref{fig9}(e), the harmonic emission is dominated by the short path. The calculated HHG yields first increase rapidly as the laser intensity changes from $0.5\times10^{14}\ \mathrm{W/cm}^2$ to $2\times10^{14}\ \mathrm{W/cm}^2$. In this range, the intensity modulation is unapparent. However, when the laser intensity is over $2\times10^{14}\ \mathrm{W/cm}^2$, obvious modulation structures emerge (see the illustration). For different harmonic orders (here, H15, H17, and H19 are selected as examples), almost the same modulation period of about $0.3\sim0.4\times10^{14}\ \mathrm{W/cm}^2$ is shown. Note that the QPI does not work in this case and the modulation is mainly caused by the transient phase matching of the short path. Similar results can also be found in Fig. \ref{fig9}(f), where the long path contribution is dominant. Here, it is worth mentioning that in Fig. \ref{fig9}(e), one can also see the modulations in the intensity from $0.5\times10^{14}\ \mathrm{W/cm}^2$ to $2\times10^{14}\ \mathrm{W/cm}^2$. This is mainly due to the imperfect selection of the long path. These results show that the intensity-dependent interference fringe caused by the macroscopic phase-matching depends less sensitively on the harmonic order. This provides a criteria to differentiate the interference caused by QPI and macroscopic phase-matching effects. Recall that even though short and long paths coexist and the modulation depth is indeed smaller as comparing Fig. \ref{fig7}(b) with \ref{fig7}(a), the modulation period depends insensitively on the harmonic order and dominant quantum path. Therefore, the observed intensity-dependent modulations in our experiment can be mainly attributed to the macroscopic phase-matching of each individual quantum path, which probably is dominant over the QPI.

\begin{figure}[htb]
\centerline{
\includegraphics[width=8.5cm]{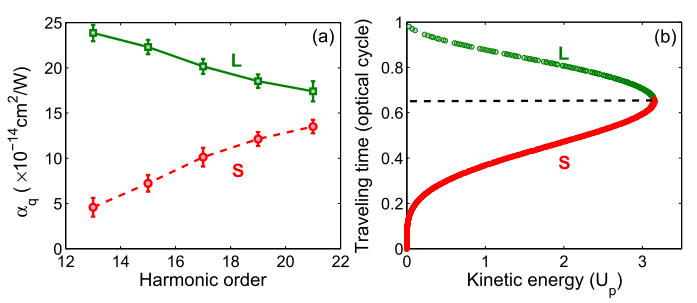}}
 \caption{\label{fig10}(Color online) (a) Experimentally retrieved dipole phase coefficients of the short and long paths. (b) Calculated traveling times of the short and long paths.
 }
\end{figure}

\section{Retrieving the atomic dipole phase and temporal chirp}

In section \uppercase\expandafter{\romannumeral2}, we have shown that the spectrum splitting and frequency shift are mainly induced by the transient phase-matching due to the variation of the atomic dipole phases of different quantum paths in time. In turn, we can retrieve the atomic dipole phase via the intensity-dependent frequency shift. Moreover, the spatially and spectrally resolved HHG provides the information for both the short and long paths simultaneously. According to Eq. (2), we know that when the laser intensity $I_0$ reaches the phase matching intensity $I_m$, the frequency shift of the $q$th harmonic approaches to zero. This allows us to directly read out the $I_m$ (where the frequency shift is minimum) for each harmonic from the measured intensity-dependent harmonic spectra.
As shown in Fig. \ref{fig2}(d), for the short paths of H13$-$H19 (right branches), the transient phase-matching intensities can be directly read out at the minimum frequency shift, which are $1.4\times10^{14}\ \mathrm{W/cm}^2$ (1.59 mJ), $1.34\times10^{14}\ \mathrm{W/cm}^2$ (1.52 mJ), $1.3\times10^{14}\ \mathrm{W/cm}^2$ (1.47 mJ), and $1.25\times10^{14}\ \mathrm{W/cm}^2$ (1.41 mJ), respectively. Then one can retrieve the phase coefficient for each harmonic ($\alpha_{13}^{S}=4.6\times10^{-14}\ \mathrm{cm^2/W}$, $\alpha_{15}^{S}=7.2\times10^{-14}\ \mathrm{cm^2/W}$, $\alpha_{17}^{S}=10\times10^{-14}\ \mathrm{cm^2/W}$, and $\alpha_{19}^{S}=12\times10^{-14}\ \mathrm{cm^2/W}$) by fitting the intensity-dependent red-shift value according to Eq. (2). In principle, the phase coefficient $\alpha_q$ of the long quantum path can be retrieved in the same way. However, the transient phase-matching intensity of the long path is about one-order of magnitude smaller than that of the short one, i.e., $<0.4\times10^{14}\ \mathrm{W/cm}^2$, where the harmonic signal and cutoff energy are very low. Hence it makes it difficult to directly obtain the $I_m$ as the short path. Then, we retrieve the phase coefficient by seeking the best fit value by setting $\alpha_q$ in the range of [15, 25] $\times10^{-14}\ \mathrm{cm^2/W}$ and $I_m$ in the range of [0.1, 0.4] $\times10^{14}\ \mathrm{W/cm}^2$. Finally, the optimal values of $I_m$ of the long paths for H13$-$H19 are determined as $0.18\times10^{14}\ \mathrm{W/cm}^2$ (0.2 mJ), $0.23\times10^{14}\ \mathrm{W/cm}^2$ (0.26 mJ), $0.27\times10^{14}\ \mathrm{W/cm}^2$ (0.3 mJ), and $0.31\times10^{14}\ \mathrm{W/cm}^2$ (0.35 mJ). Corresponding retrieved phase coefficients are $\alpha_{13}^{L}=24\times10^{-14}\ \mathrm{cm^2/W}$, $\alpha_{15}^{L}=22.6\times10^{-14}\ \mathrm{cm^2/W}$, $\alpha_{17}^{L}=20\times10^{-14}\ \mathrm{cm^2/W}$, and $\alpha_{19}^{L}=18.5\times10^{-14}\ \mathrm{cm^2/W}$, respectively.
Figure \ref{fig10}(a) plots the retrieved $\alpha_q$ of the short (circles) and long (squares) paths as a function of harmonic orders. Clearly, $\alpha_q$ increases as harmonic order increases for the short path. In contrast, $\alpha_q$ decreases for the long path, which is in agreement with those extrapolated from the chirp measurements \cite{m1,m2} and theoretical calculations \cite{t1,t2}. According to the SFA model, the phase coefficients are directly linked to the excursion times of quantum paths in terms of $\Phi_{q}^{j}\approx-U_p\tau_{q}^{j}\approx-\alpha_{q}^{j}I$.
For comparison, the calculated excursion times of the short and long paths are also presented in Fig. \ref{fig10}(b). The experimental curves show very similar trends to the theoretical calculations as the harmonic order increases: the short path shows a positive temporal chirp while the long path shows a negative one.

\section{Conclusion}
In conclusion, we have simultaneously observed the spectrum splitting, frequency shift and intensity-dependent interference both for the short
and long paths in experiment.
It is shown that each
individual high-order harmonic is gradually split into two branches in the spatial and spectral domains as the laser intensity increases.
By adjusting the focusing position and laser
intensity, each branch shows distinct spatial profiles, either
blue- or red-frequency shift. The HHG yield also shows an intensity-dependent interference fringe.
The simulations indicate that the interference fringe due to QPI depends on the harmonic order. In contrast, the interference fringe due to the transient phase matching of individual quantum path depends less sensitively on the harmonic order. This provides a robust criteria to differentiate these two interference mechanisms.
Moreover, the atomic dipole phase is encoded in the intensity-dependent frequency shift, from which we retrieve the atomic dipole phases and reveal the temporal chirps both for the short and long paths.

The heart underlying the rich spatiotemporal and spectral features is the atomic dipole phase $\Phi_q=-\alpha_qI(t)$ associated with different quantum paths.
The different phase coefficients lead to spatially resolved harmonic signals for short and long paths. The evolution of the intensity-dependent dipole phase
in the time domain leads to transient phase matching, which results in spectral splitting, intensity-dependent frequency shift and interference of HHG yields.
The frequency shift enables us to measure the phase coefficient $\alpha_q$, which depends on the harmonic order. In turn the order-dependent phase coefficients
reveal the positive and negative temporal chirps of the short and long paths, i.e., the time-frequency property of HHG in the sub-cycle of the laser field.

Previous attosecond spectroscopy techniques are mainly concentrated on the spectral features of harmonic spectra or the temporal features of the generated attosecond pulses. On the other side, as coherent lights, high harmonics contain other intrinsic features, such as the spatial profiles, interference fringes, etc..  A few recent investigations have proposed and demonstrated that these new features can shed new light on the molecular internal dynamics \cite{azc} and {\it in situ} measurement of isolated attosecond pulses in space and time \cite{npy}. Our work provides an informative mapping of the spatiotemporal and spectral features of quantum paths in HHG. These observed HHG features can serve as promising tools for attosecond spectroscopy in both spectral and spatiotemporal dimensions.

\section*{Acknowledgement}

This work was supported by the 973 Program of China under Grant No. 2011CB808103 and the National Natural Science Foundation of China under Grants
No. 61275126, 11234004, and 0204012208. Numerical simulations presented in
this paper were carried out using the High Performance Computing experimental testbed in SCTS/CGCL (see
http://grid.hust.edu.cn/hpcc).

\end{document}